\documentclass[prl,twocolumn,lengthcheck,superscriptaddress,letterpaper,nofootinbib,nopreprintnumbers,showpacs]{revtex4}

\usepackage{color}
\usepackage{graphicx}
\usepackage{float}
\usepackage{amsmath}
\usepackage{acronym}
\usepackage{multirow}

\usepackage{color}

\newcommand{\beq}{\begin{equation}}
\newcommand{\eeq}{\end{equation}}
\newcommand{\bea}{\begin{eqnarray}} 
\newcommand{\eea}{\end{eqnarray}}
\newcommand{\ba}{\begin{array}}
\newcommand{\ea}{\end{array}}

\newcommand{\mytextrm}[1]{{}}

\def\ltsima{$\; \buildrel < \over \sim \;$}
\def\simlt{\lower.5ex\hbox{\ltsima}}
\def\gtsima{$\; \buildrel > \over \sim \;$}
\def\simgt{\lower.5ex\hbox{\gtsima}}

\begin{document}

\title{Observing the dynamics of super-massive black hole binaries with Pulsar Timing Arrays}


\author{C. M. F. Mingarelli}
\affiliation{School of Physics and Astronomy, University of Birmingham, Edgbaston, Birmingham B15 2TT, UK}

\author{K. Grover}
\affiliation{School of Physics and Astronomy, University of Birmingham, Edgbaston, Birmingham B15 2TT, UK}

\author{T. Sidery}
\affiliation{School of Physics and Astronomy, University of Birmingham, Edgbaston, Birmingham B15 2TT, UK}

\author{R. J. E. Smith}
\affiliation{School of Physics and Astronomy, University of Birmingham, Edgbaston, Birmingham B15 2TT, UK}
\affiliation{Perimeter Institute for Theoretical Physics, Waterloo, Ontario N2L 2Y5, Canada}
\author{A. Vecchio}
\affiliation{School of Physics and Astronomy, University of Birmingham, Edgbaston, Birmingham B15 2TT, UK}

\date\today

\begin{abstract}
Pulsar Timing Arrays are a prime tool to study unexplored astrophysical regimes with gravitational waves. Here we show that the detection of gravitational radiation from individually resolvable super-massive black hole binary systems can yield direct information about the masses and spins of the black holes, provided that the gravitational-wave induced timing fluctuations both at the pulsar and at the Earth are detected. This in turn provides a map of the non-linear dynamics of the gravitational field and a new avenue to tackle open problems in astrophysics connected to the formation and evolution of super-massive black holes. We discuss the potential, the challenges and the limitations of these observations.
\end{abstract}


\maketitle

\textit{Introduction---}
Gravitational waves (GWs) provide a new means for studying black holes and addressing open questions in astrophysics and fundamental physics: from their formation, evolution and demographics, to the assembly history of galactic structures and the dynamical behaviour of gravitational fields in the strong non-linear regime. Specifically, GW observations through a network of radio pulsars used as ultra-stable clocks -- Pulsar Timing Arrays (PTAs)~\cite{EstabrookWahlquist:1975, Sazhin:1978, Detweiler:1979} -- represent the only \emph{direct} observational avenue for the study of supermassive black hole binary (SMBHB) systems in the $\sim 10^8-  10^9 M_{\odot}$ mass range, with orbital periods between $\sim1$ month and a few years, see \emph{e.g.}~\cite{Sesana:2012, Volonteri:2010} and references therein. Ongoing observations~\cite{PPTA, EPTA, NANOGrav,IPTA} and future instruments, \emph{e.g.} the Square Kilometre Array~\cite{ska-web}, are expected to yield the necessary timing precision~\cite{VerbiestEtAl:2009, LiuEtAl:2011} to observe the diffuse GW background.
This background is likely dominated by the incoherent superposition of radiation from the cosmic population of massive black holes \cite{HellingsDowns:1983,RajagopalRomani:1995,WyitheLoeb:2003, SesanaEtAl:2004, JaffeBacker:2003, JenetEtAl:2006,SesanaVecchioColacino:2008,vanHaasterenEtAl:2011,DemorestEtAl:2012} and within it, we expect a handful of sources that are sufficiently close, massive and high-frequency to be individually resolvable ~\cite{JenetEtAl:2004, SesanaVecchioVolonteri:2008, SesanaVecchio:2010, YardleyEtAl:2010, WenEtAl:2011, LeeEtAl:2011, BabakSesana:2012, EllisJenetMcLaughlin:2012, EllisSiemensCreighton:2012}.

Massive black hole formation and evolution scenarios~\cite{VolonteriHaardtMadau:2003, KoushiappasZentner:2006, MalbonEtAl:2007, YooEtAl:2007} predict the existence of a large number of SMBHBs. Furthermore, SMBHs are expected to be (possibly rapidly) spinning~\cite{MerrittEkers:2002,HughesBlandford:2003}. In fact the dynamics of such systems -- which according to general relativity are entirely determined by the masses and spins of the black holes~\cite{MTW} -- leave a direct imprint on the emitted gravitational waveforms. From these, one could measure SMBHB masses and their distribution, yielding new insights into the assembly of galaxies and the dynamical processes in galactic nuclei~\cite{WenEtAl:2011}.
Moreover, measuring the magnitude and/or orientation of spins in SMBHBs would provide new information on the role of accretion processes~\cite{GammieEtAl:2004, VolonteriEtAl:2005, BertiVolonteri:2008, PeregoEtAl:2009, DottiEtAl:2010}. Finally, detections of SMBHBs could allow us to probe general relativistic effects in the non-linear regime in an astrophysical context not directly accessible by other means, see ~\cite{LLRtestGR} and references therein.

The observation of GWs with PTAs relies on the detection of the small deviation induced by gravitational radiation in the times of arrival (TOAs) of radio pulses from millisecond pulsars that function as ultra-stable reference clocks. This deviation, called the residual, is the difference between the expected (without GW contribution) and actual TOAs once all the other physical effects are taken into account.  The imprint of GWs on the timing residuals is the result of how the propagation of radio waves is affected by the GW-induced space-time perturbations along the travel path. It is a linear combination of the GW perturbation at the time when the radiation transits at a pulsar, the so-called ``pulsar term'', and then when it passes at the radio receiver, the ``Earth term"~\cite{EstabrookWahlquist:1975, Sazhin:1978, Detweiler:1979}. The two terms reflect the state of a GW source at two \emph{different} times of its evolution separated by $\tau \equiv(1 + \hat{\mathbf{\Omega}} \cdot \hat{\bf{p}})\,L_p\sim  3.3\times 10^3\,(1 + \hat{\mathbf{\Omega}} \cdot \hat{\bf{p}})\,(L_p/1\,\mathrm{kpc})$~yr where $\hat{\mathbf{\Omega}}$ and $\hat{\bf{p}}$ are the unit vectors that identify the GW propagation direction and the pulsar sky location at a distance $L_p$ from the Earth, respectively, see \emph{e.g.}~\cite{SesanaVecchio:2010}.  [We use geometrical units in which $G = c = 1$.] In a network (array) of pulsars all the perturbations at the Earth add coherently and therefore boost the signal-to-noise ratio (S/N) of the signal. Each pulsar term is at a slightly different frequency since the orbital period of the binary evolves over the time $\tau$. 

Measuring the key physics of SMBHBs is hampered by the short (typically $T=10$~yr) observation time compared to the typical orbital evolution timescale $f/\dot{f}  = 1.6\times 10^3 (\mathcal{M}/10^9\,M_\odot)^{-5/3} (f/50\,\mathrm{nHz})^{-8/3}$~yr of binaries that are still in the weak field adiabatic inspiral regime, with an orbital velocity $v = 0.12\,(M/10^9\,M_\odot)^{1/3}\,(f/50\,\mathrm{nHz})^{1/3}$~\cite{Peters:1964}. Here $M = m_1 + m_2$, $\mu = m_1 m_2/M$ and $\mathcal{M} = M^{2/5} \mu^{3/5}$ are the total, reduced and chirp mass, respectively, of a binary with component masses $m_{1,2}$, and $f$ is the GW emission frequency at the leading quadrupole order. The chirp mass determines the frequency evolution at the leading Newtonian order. In the post-Newtonian (pN) expansion of the binary evolution~\cite{Blanchet:2006} in terms of $v\ll 1$, the second mass parameter enters at p$^1$N order (${\mathcal O}(v^2)$ correction); spins contribute at p$^{1.5}$N order and above  (${\mathcal O}(v^3)$) causing the orbital plane to precess through spin-orbit coupling, at leading order. These contributions are therefore seemingly out of observational reach.

The GW effect at the pulsar -- the pulsar term -- may be detectable in future surveys, and for selected pulsars their distance could be determined to sub-parsec precision~\cite{LeeEtAl:2011,SmitsEtAl:2011, DellerEtAl:2008}. If this is indeed the case, it opens the opportunity to coherently connect the signal observed at the Earth and at 
pulsars, therefore providing snapshots of the binary evolution over $\sim 10^3$ yr. These observations would drastically change the ability to infer SMBHB dynamics, and study the relevant astrophysical process and fundamental physics.

In this \emph{Letter} we show that for SMBHBs at the high end of the mass and frequency spectrum observable by PTAs, say $m_{1,2} = 10^9\,M_\odot$ and $f = 10^{-7}$ Hz, the observations  of a source still in the weak-field regime become sensitive to post-Newtonian contributions up to p$^{1.5}$N, including spin-orbit effects, if both the pulsar and Earth term can be detected. This in principle enables the measurement of the two mass parameters and a combination of the spin's magnitude and relative orientation.
We also show that the Earth-term only can still be sensitive to spin-orbit coupling due to geometrical effects produced by precession.  We discuss the key factors that enable these measurements, and future observational prospects and limitations.

\textit{Signals from SMBHBs---} Consider a radio pulsar emitting radio pulses at frequency $\nu_{0}$ in the source rest-frame. GWs modify the rate at which the radio signals are received at the Earth~\cite{EstabrookWahlquist:1975, Sazhin:1978, Detweiler:1979}, inducing a relative frequency shift  ${\delta\nu(t)}/{\nu_0} = h(t - \tau) - h(t)$, where $h(t)$ is the GW strain.The quantities that are actually produced at the end of the data reduction process of a PTA are the timing residuals, $\int dt'\,{\delta\nu(t')}/{\nu_0}$, although without loss of generality, we will base the discussion on $h(t)$. The perturbation induced by GWs is repeated twice, and carries information about the source at time $t$, the ``Earth term", and at past time $t - \tau$, the ``pulsar term". 

We model the radiation from a SMBHB using the so-called restricted pN approximation, in which pN corrections are included only in the phase and the amplitude is retained at the leading Newtonian order, but we include the leading order modulation effects produced by spin-orbit coupling. The strain is given by 
\beq\label{eq:zed}
h(t)= - A_\mathrm{gw}(t) A_\mathrm{p}(t) \cos[\Phi(t)+\varphi_\mathrm{p}(t)+\varphi_\mathrm{T}(t)]\,,
\eeq
where $A_\mathrm{gw}(t) = 2 [\pi f(t)]^{2/3} \mathcal{M}^{5/3}/D$ is the Newtonian order GW amplitude, $\Phi(t)$ is the GW phase, see \emph{e.g.} Eq.~(232, 234) in~\cite{Blanchet:2006} and Eq.~(8.4) in~\cite{BlanchetBuonannoFaye:2006}, and $D$ is the distance to the GW source. $A_p(t)$ and $\varphi_p(t)$ are the time-dependent polarisation amplitude and phase and $\varphi_\mathrm{T}(t)$ is an additional phase term, analogous to Thomas precession, see Eq.~(29) in~\cite{ApostolatosEtAl:1994}.

The physical parameters leave different observational signatures in the GW strain $h(t)$ and are therefore found in the TOA residuals. At the leading Newtonian order, $\mathcal{M}$ drives the frequency and therefore the phase $\Phi(t)$ evolution, with the second independent mass parameter entering from the p$^1$N onwards. SMBHs are believed to be rapidly spinning, and the spins are responsible for three distinctive imprints in the waveform: (i) they alter the phase evolution through spin-orbit coupling and spin-spin coupling at p$^{1.5}$N and p$^{2}$N order, respectively~\cite{KidderEtAl:1993},
(ii) they cause the orbital plane to precess due to (at lowest order) spin-orbit coupling and therefore induce amplitude and phase modulations in the waveform through $A_p(t)$ and $\varphi_p(t)$; and (iii) through
orbital precession they introduce an additional secular contribution $\varphi_\mathrm{T}(t)$ to the waveform phase. Astrophysically we expect PTAs to detect SMBHBs of comparable component masses~\cite{SesanaVecchio:2010}. We therefore model the spin-orbit precession using the \emph{simple precession approximation}~\cite{ApostolatosEtAl:1994}, which formally applies when $m_1 = m_2$, or when one of the two spins is negligible with respect to the other. Let ${\mathbf S}_{1,2}$ and $\mathbf{L}$ be the black holes' spins and the orbital angular momentum, respectively. Then both ${\mathbf S} = {\mathbf S}_1 + {\mathbf S}_2$ and $\mathbf{L}$ precess around the (essentially) constant direction of the total angular momentum, $ \mathbf{{J}} = {\mathbf S} + \mathbf{L}$, at the same rate ${d\alpha}/{dt}=\pi^2\left(2+{3m_2}/({2m_1})\right)({|{\mathbf{ L}}+{\mathbf {S}}|})f^2(t)/M$~\cite{ApostolatosEtAl:1994}, where $\alpha$ is the precession angle, while preserving the angle of the precession cone, $\lambda_L$, see Fig. 4 of Ref.~\cite{ApostolatosEtAl:1994}. This approximation is adequate to conceptually explore these effects, however in the case of real observations, one will need to consider the exact expressions~\cite{Kidder:1995}.

The detection and particularly the measurement of the aforementioned parameters relies on coherently matching the signal with a template that faithfully reproduces its amplitude and, importantly, its phase evolution. We 
 therefore consider the contribution to the total number of wave cycles a proxy for the significance of a specific parameter. Individual terms that contribute $\sim 1$ GW cycle or more mean that the effect is in principle detectable, hence one can infer information about the associated parameter(s).
We show that information about the parameters can only be inferred for SMBHBs at the high end of the mass spectrum and PTA observational frequency range. Having a sufficiently high-mass and high-frequency GW source is also essential to ensure sufficient frequency evolution over the time $\tau$, so that the Earth and pulsar term are clearly separated in frequency space cf. Table~\ref{t:gw_cycles}. We therefore take fiducial source parameters of $m_1 = m_2 = 10^9\,M_\odot$, frequency at the Earth at the beginning of the observation $f_E = 10^{-7}$ Hz and an observational time $T = 10$ years to illustrate the main results. We provide scaling relations as a function of the relevant quantities, allowing the reader to rescale the results for different astrophysical and/or observational values.

\textit{Observations using the Earth-term only ---} {We start by considering analyses that rely only on the Earth-term contribution to the residuals, as done in Ref.~\cite{LommenBacker:2001, YardleyEtAl:2010}. The case of a coherent analysis based both on the Earth- and pulsar-term, introduced in Ref.~\cite{JenetEtAl:2004}, is discussed later in this Letter.
Table~\ref{t:gw_cycles} shows that, in general, the frequency change over 10 yr is small compared to the frequency bin width, $3.2(10\,\mathrm{yr}/T)$~nHz~\cite{LeeEtAl:2011, SesanaVecchio:2010}. The observed signal is effectively monochromatic, making the dynamics of the system impossible to infer. However, the presence of spins affects the waveform not only through the GW phase evolution, but also via the modulations of $A_\mathrm{p}(t)$ and $\varphi_\mathrm{p}(t)$ that are periodic over the precession period, \emph{and} also introduces the secular contribution $\varphi_\mathrm{T}(t)$. For $m_{1,2} = 10^9\,M_\odot$ and $f_E = 10^{-7}$~Hz
the orbital angular momentum precesses by $\Delta\alpha = 2$~rad (for dimensionless spin parameter $a \equiv S/M^2 = 0.1$) and $\Delta\alpha = 3$~rad (for $a = 0.98$), and therefore the additional modulation effect on $A_\mathrm{p}(t)$ and $\varphi_\mathrm{p}(t)$ is small, and likely undetectable. However, the overall change of $\varphi_\mathrm{T}(t)$ 
over 10~yrs could be appreciable: the average contribution for each precession cycle of this additional phase term is $\langle \Delta \varphi_\mathrm{T} \rangle = 4 \pi$ or $4 \pi\, (1-\cos\lambda_L)$, depending on whether $\hat{\mathbf{\Omega}}$ lies inside or outside the precession cone, respectively ~\cite{ApostolatosEtAl:1994}. If $\hat{\mathbf{\Omega}}$ lies inside the precession cone, and given that the observation will cover between a third and a half of a full precession cycle, then $\langle\Delta\varphi_\mathrm{T}\rangle\sim \pi$, which could surely indicate the presence of spins. On the other hand, the precession cone will be small in general since 
$|S/L| \sim a\,v\,(M/\mu) \simeq 0.1\,a\,(M/\mu)\,(M/10^9\,M_\odot)^{1/3}\,(f/100\,\mathrm{nHz})^{1/3}$, therefore the likelihood of $\hat{\mathbf{\Omega}}$ lying inside the precession cone is small, assuming an isotropic distribution and orientation of sources. In this case the Thomas precession contribution (per precession cycle) is suppressed by a factor 
$(1-\cos\lambda_L ) \simeq \lambda_L^2/2 \sim 5\times 10^{-3}\,a^2\,(M/\mu)^2\,(M/10^9\,M_\odot)^{2/3}\,(f/100\,\mathrm{nHz})^{2/3}$,
which will produce a negligible contribution $\Delta\varphi_\mathrm{T}(t) \ll 1$. However unlikely, spins may still introduce observable effects that need to be taken into account.

\begin{table*}[htdp]
\begin{center}
\begin{tabular}{|cccccc|cccccc|}
\hline
$m_1(M_\odot)	$				& 		$m_2(M_\odot)$ 		&	 $f_E$(nHz)		&	$(v/c\times 10^{-2})$		&	timespan	& 	$\Delta f$ (nHz) 	&	Total		& Newtonian			&	p$^1$N		& 	p$^{1.5}$N	&	spin-orbit$/\beta$	&	p$^{2}$N 		\\
\hline
\multirow{4}{*}{$10^9$} 	& \multirow{4}{*}{$10^9$} &\multirow{2}{*}{100}&	14.6				& 	10 yr			&	3.22				&	 32.1		&	31.7			& 0.9 		&  -0.7	 	& 	0.06		&  	0.04		 	\\
					&					& 				&	9.6				& 	-1 kpc		&	71.2				&	4305.1	&	4267.8		&77.3		&  -45.8 		& 	3.6		& 	2.2		 	\\
					&					&\multirow{2}{*}{50}	&	11.6				& 	10 yr			&	 0.24				&	15.8		&	15.7			& 0.3			& -0.2		&  	0.01		& 	$< 0.01$		\\
					&					& 				&	9.4				& 	-1 kpc		&	 23.1				&	3533.1	&	3504.8		& 53.5		& -28.7		&  	2.3		& 	1.2			\\
					
\hline
\multirow{4}{*}{$10^8$} 	& \multirow{4}{*}{$10^8$} &\multirow{2}{*}{100}&	6.8				& 	10 yr			&	0.07				&	 31.6		&	31.4			& 0.2			&   -0.07		& 	$< 0.01$	&   	$< 0.01$		\\
					&					& 				&	6.4				& 	-1 kpc 		&	15.8 				&	9396.3	&	9355.7		&58.3		&  -19.9		& 	1.6		&  	0.5			\\
					&					&\multirow{2}{*}{50} &	5.4				& 	10 yr			&	 0.005			&	15.8		&	15.7			& 0.06		&  -0.02		&  	$< 0.01$	& 	$< 0.01$		\\
					&					& 				&	5.3				& 	-1 kpc		&	1.62		 		&	5061.4	&	5045.8		& 20.8		&  -5.8		& 	0.5	 	& 	0.1			\\
					
\hline
\end{tabular}
\end{center}
\caption {Frequency change $\Delta f$, total number of GW cycles and individual contributions from the leading order terms in the pN expansion over the two relevant time scales -- a 10 yr period starting at the Earth 
and the time period $L_p (1 + \hat{\mathbf{\Omega}} \cdot \hat{\bf{p}})$ between the Earth and pulsar term (hence the negative sign) -- for selected values of $m_{1,2}$ and $f_E$.}
%:  one starting at $f_E$ and covering 10 yrs, and the other starting at $L_p (1 + \hat{\mathbf{\Omega}} \cdot \hat{\bf{p}})$ in the past (hence the negative sign) 
%ending at $f_E+10$~yrs and the other starting at 1 kpc $(\sim 3,300$~yrs) in the past, hence the negative sign, and ending at $f_E$. This has been done for selected values of $m_{1,2}$ and $f_E$}}
\label{t:gw_cycles}
\end{table*}

\textit{Measuring SMBHB evolution using the Earth and pulsar term---} 
With more sensitive observations and the increasing possibility of precisely determining $L_p$ see \emph{e.g.}~\cite{SmitsEtAl:2011}, the prospect of also observing the contribution from the pulsar-term from one or more pulsars becomes more realistic. We show below that \emph{if} at least one of the pulsar terms can be observed together with the Earth-term, this opens opportunities to study the dynamical evolution of SMBHBs and, in principle, to measure their masses and spins. This is a straightforward consequence of the fact that PTAs become sensitive to $\sim 10^3$~yr of SMBHB evolutionary history, in ``snippets'' of length $T \ll L_p$ that can be coherently concatenated. 

The signal from each pulsar term will be at a S/N which is significantly smaller than the Earth-term by a factor $\sim \sqrt{N_p}$, where $N_p$ is the number of pulsars that effectively contribute to the S/N of the array. For example, if the Earth-term yields an S/N of $\sim36 \sqrt{N_p/20}$, then each individual pulsar term would give an S/N $\sim 8$. The possibility of coherently connecting the Earth-term signal with each pulsar term becomes therefore a question of S/N, prior information about the pulsar-Earth baseline and how accurately the SMBHB location in the sky can be reconstructed, as part of a ``global fit", \emph{e.g.}~\cite{LeeEtAl:2011}.
Assuming for simplicity that the uncertainties on $L_p$ and $\hat{\mathbf{\Omega}}$ are uncorrelated, this requires that the distance to the pulsar and the location of the GW source are known with errors $\simlt 0.01 (100\,\mathrm{nHz}/f)$~pc and $\simlt 3 (100\,\mathrm{nHz}/f) (1\,\mathrm{kpc}/L_p)$~arcsec, respectively. These are very stringent constraints~\cite{SmitsEtAl:2011, SesanaVecchio:2010, BabakSesana:2012}, and a detailed analysis is needed in order to assess the feasibility of reaching this precision. Clearly if an electromagnetic counterpart to the GW source were to be found~\cite{SesanaEtAl:2012, TanakaEtAl:2012}, it would enable the identification of the source location in the sky, making the latter constraint unnecessary.

We can now consider the contribution from the different terms in the pN expansion to the total number of cycles in observations that cover the GW source evolution over the time $\tau$
that are encoded in the simultaneous analysis of the Earth \emph{and} pulsar terms.  The results are summarised in Table~\ref{t:gw_cycles}, for selected values of $m_{1,2}$, and $f_E$ and for a fiducial value $\tau = 1\,\mathrm{kpc}$. The wavecycle contributions from the spin-orbit parameter are normalised to $\beta = (1/12) \sum_{i=1}^2 \left[113 (m_i/M)^2 + 75\eta\right] \hat{\mathbf{L}} \cdot \hat{\mathbf{S}}_i$, which has a maximum value of 7.8. Contributions from the p$^2$N order spin-spin terms are negligible. The results clearly show that despite the fact that the source is in the weak field regime the extended Earth-pulsar baseline requires the p$^{1.5}$N, and in some rare cases the p$^2$N contribution, to accurately (\emph{i.e.} within $\sim 1$ GW cycle) reproduce the full phase evolution.}

For $m_{1,2} = 10^9\,M_\odot$ and $f_E=10^{-7}$ Hz there is a total of 4305 GW cycles over a 1 kpc light travel time evolution,
 with the majority (4267) accounted for by the leading order Newtonian term, providing information about the chirp mass, and tens of cycles due to the p$^{1}$N  and p$^{1.5}$N terms (77 and 45, respectively), that provide information about a second independent mass parameter. Spins contribute to phasing at p$^{1.5}$N with $\sim 3\beta$ cycles. Therefore their total contribution is smaller than the p$^{1.5}$N mass contribution by a factor between a few and $\sim 10$ . The additional Thomas precession phase contribution may become comparable to the p$^{1}$N mass contribution in some cases. In fact, for $a= 0.1 (0.98)$ the binary undergoes $24\, (34)$ precession cycles. This corresponds to a total Thomas precession phase contribution of 306 (426) rad  if $\hat{\mathbf{\Omega}}$ lies outside the precession cone.

The modulations of $A_\mathrm{p}(t)$ and $\varphi_\mathrm{p}(t)$ are characterised by a small $\lambda_\mathrm{L}$, because for most of the inspiral $S \ll L$, and are likely to leave a smaller imprint on the waveform than those discussed so far. We can indeed estimate the importance of this effect for the most favourable parameter combinations. The value of 
$\varphi_\mathrm{p}(t)$ oscillates over time with an amplitude which depends on the time to coalescence, $\mathbf{S}$, $\mathbf{L}$, $\mathbf{\hat{\Omega}}$ and $\mathbf{\hat{p}}$. We choose the orientation of $\bf{\hat{S}}$ such that $\lambda_L$ is maximised, and we vary $\mathbf{\hat{\Omega}}$ and $\mathbf{\hat{p}}$, each of which is drawn from a uniform distribution on the two-sphere.
 
In Figure 1 we show that for rapidly spinning ($a=0.98$) SMBHBs this effect could introduce modulations larger than $\pi/2$ in $\varphi_\mathrm{p}(t)$ over 30\% of the parameter space of possible $\mathbf{\hat{\Omega}}$ and $\mathbf{\hat{p}}$ geometries. The amplitude would correspondingly change over the same portion of the parameter space by at most 60\% with respect to its unmodulated value. Since these effects are modulated, they will not be easily identifiable. 

\textit{Conclusions---} We have established that the coherent observation of both the Earth and pulsar term provides
information about the dynamical evolution of a GW source. The question now is whether they can be unambiguously identified.
A rigorous analysis would require extensive simulations based on the actual analysis of synthetic data sets. We can however gain the key information with a much simpler order of magnitude calculation. The phase (or number of cycles) error scales as $\sim 1/(\mathrm{S/N})$. Assuming S/N $\sim 40$ means that the total number of wave cycles over the Earth-pulsar baseline can be determined with an error $\sim 4300/40\sim 100$ wave cycles. This is comparable to the p$^1$N contribution to the GW phase and, in very favourable circumstances, to the Thomas precession phase contribution, and larger by a factor of a few or more than all the other contributions. It may therefore be possible to measure the chirp mass and, say, the symmetric mass ratio of a SMBHB, and possibly a combination of the spin parameters. Effects due to the p$^{1.5}$N and higher phase terms are likely to remain unobservable, as well as amplitude and phase modulations. Correlations between the parameters, in particular masses and spins, will further degrade the measurements. The details will depend on the actual S/N of the observations, the GW source parameters, and the accuracy with which the source location and the pulsar distance can be determined. We plan to explore these issues in detail in a future study.

\begin{figure}
\includegraphics[scale=0.45]{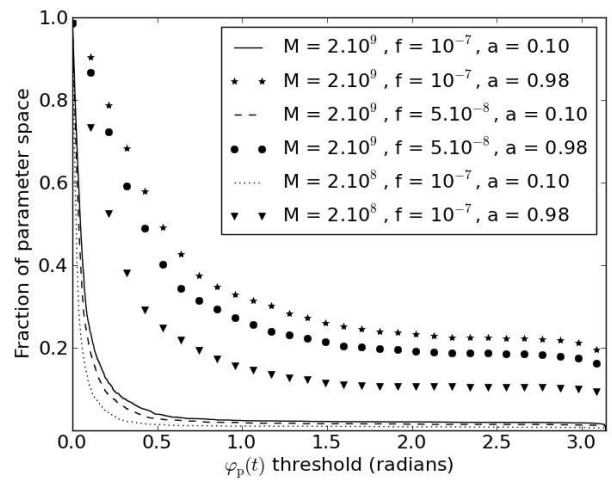}
\label{fig:phase}
\caption{The fraction of parameter space in $\hat{\mathbf{\Omega}}$ and $\hat{\bf{p}}$ for which the maximum excursion of $\varphi_\mathrm{p}$ over the time $L_p (1 + \hat{\mathbf{\Omega}} \cdot \hat{\bf{p}})$ for $L_p = 1\,\mathrm{kpc}$ exceeds a certain value, shown on the horizontal axis. Several values of $m_{1,2}$, $a$ and $f_\mathrm{E}$ are considered (see legend).}
\end{figure}

{\it Acknowledgements---}  We would like to thank the referees for their suggestions and comments. Furthermore, we would like to thank I.~Mandel,  K.~J.~Lee, A.~Sesana , J.~Verbiest and our colleagues of the European Pulsar Timing Array Collaboration for many useful discussions and comments. This work has been supported by the UK Science and Technology Facilities Council. RJES acknowledges the support of a Perimeter Institute Visiting Graduate Fellowship.  CMFM acknowledges the support of the Royal Astronomical Society and the Institute of Physics.

\end{document}